\documentclass[aps,notitlepage,prb,floatfix,showpacs,superscriptaddress]{revtex4-1}

\usepackage{amsmath}
\usepackage{amssymb}
\usepackage{amsfonts}
\usepackage[pdftex]{graphicx}
\usepackage{units}
\usepackage[usenames]{color}
\usepackage{dcolumn}
\usepackage{bm}
\usepackage{float}
\usepackage[utf8]{inputenc}
\usepackage[colorlinks=true,citecolor=blue,linkcolor=blue]{hyperref}

\usepackage{epstopdf}

\definecolor{cream}{RGB}{222,217,201}

\begin{document}

\title{Effects of hole doping on the electronic and optical properties of transparent conducting copper iodide}

\author{Michael Seifert}
\affiliation{Institut für Festkörpertheorie und -optik, Friedrich-Schiller-Universität Jena, Max-Wien-Platz 1, 07743 Jena, Germany}
\affiliation{European Theoretical Spectroscopy Facility}
\author{Miguel A. L. Marques}
\affiliation{Institut für Physik, Martin-Luther-Universität Halle-Wittenberg, 06120 Halle/Saale, Germany}
\affiliation{European Theoretical Spectroscopy Facility}
\author{Silvana Botti}
\affiliation{Institut für Festkörpertheorie und -optik, Friedrich-Schiller-Universität Jena, Max-Wien-Platz 1, 07743 Jena, Germany}
\affiliation{European Theoretical Spectroscopy Facility}
\email{silvana.botti@uni-jena.de}

\begin{abstract}
Zincblende copper iodide has been attracting growing interest as p-type semiconductor for applications in transparent electronics and transparent thermoelectrics. A key step towards technological applications is the possibility to enhance copper iodide (CuI) conductivity by doping without deteriorating transparency. A recent high-throughput computational study revealed that chalcogen substitutions on iodine sites can act as shallow acceptors. Following computational predictions, doping by oxygen, sulfur and selenium substitutions on iodine sites has recently been realized in the laboratory, showing however that few experimental challenges have still to be tackled on the way to technological applications.  We investigate here by means of {\it ab initio} calculations the effect of such substitutions on the electronic and optical properties of CuI. Our results suggest that sulfur and selenium doping are the best candidates to obtain a controllable increase of hole concentrations, while preserving at the same time transparency in the visible and high hole mobility. 
\end{abstract}

\maketitle

\section*{Introduction}

Materials displaying at the same time high electronic conductivity and optical transparency in the visible part of the electromagnetic spectrum will be crucial for the development of future transparent electronics. This new type of devices could enable many innovative applications, such as transparent thin-film transistors~\cite{Liu_2018}, transparent electrodes~\cite{Granqvist_1993}, solar windows~\cite{Yang_2017}, or electrochromic displays~\cite{Kateb_2016}. As a matter of fact, the market of transparent displays is expected to amount to \$87.2 billion in 2025 \cite{AoLiu_2021}. High-performing n-type transparent conducting materials, like zinc oxide~\cite{Han_2016} or indium-tin oxide \cite{O'Dwyer_2009, Sakamoto_2018}, are already well established for commercial applications. However, the quality of p-type transparent thin-films~\cite{AoLiu_2021, Hu_2020, AoLiu_2020} needs to be significantly improved as existing materials still do not exhibit high enough transparency or electric conductivity.

Recently, CuI has attracted an increasing attention \cite{Grundmann_2013} thanks to its remarkable electronic properties that make it a promising p-type transparent conductor. The room-temperature zincblende $\gamma$-phase was proved to possess at the same time a large band gap of 3.1~eV \cite{Grundmann_2013, Krueger_2018} and high hole mobilities of $\mu>40$~$\text{cm}^2\text{V}^{-1}\text{s}^{-1}$ ~\cite{Yang_2017}. The temperature-dependent dielectric function of zincblende CuI has been thoroughly characterized \cite{Krueger_2018} and laser emission was recently demonstrated in CuI-based microwire cavities \cite{Wille_2017}, providing opportunities for applications in compact integrated optoelectronic circuits. 
Additionally, CuI was demonstrated to be compatible with several n-type materials~\cite{Yang_2016, Schein_2013, Ding_2012, Lee_2021}. 

The first CuI devices have already shown great potential. For example, the rectification ratios of transparent diodes of p-CuI and n-ZnO~\cite{Yang_2016} were reported to be 2~orders of magnitude higher than other diodes made of disordered phases. In perovskite solar cells, CuI thin films were successfully implemented as hole collection layers~\cite{Christians_2014}. Promising results regarding power conversion efficiency and stability are also reported \cite{Yu_2018, Matondo_2021}.  Due to the strong phonon scattering, CuI is also a promising component for thermoelectric devices \cite{AoLiu_2021}. With a figure of merit of $ZT=0.21$ at 300~K, CuI films are the best known transparent p-type thermoelectric material~\cite{Yang_2017}. Ideas of using CuI-based body-heat-driven wearable electronics, thermoelectric windows and on-chip cooling were also put forward \cite{Yang_2017, AoLiu_2021}.  CuI-based thin film transistors and displays were already developed \cite{Yang_2016, Schein_2013, Ding_2012, Lee_2021}, although there remain open issues due to poor current modulations at high hole densities~\cite{AoLiu_2021}. CuI can also be used in blue and UV LEDs \cite{Ahn_2016, Baek_2020}, and might replace GaN in the future \cite{AoLiu_2021}. Also applications as UV photodetector have been discussed \cite{Liu_2016, Yamada_2019}.

A decisive step to enable industrial applications is to control the charge carrier concentration and therefore numerous properties~\cite{Dahliah_2021} by doping the crystal of CuI. In an exhaustive high-throughput study by Graužinytė \textit{et al.}, that explored p- and n-type doping of CuI using all elements of the periodic table, chalcogen elements were highlighted as potential shallow acceptors due to thermodynamic transition levels close to the valence favoring hole generating charge states~\cite{Grauzinyte_2019}. Sulfur (S) and selenium (Se) were furthermore highlighted to display lower formation energies and higher thermal equilibrium charged defect concentrations than  copper vacancies~\cite{Grauzinyte_2019}, the natural occurring  acceptor defects of CuI~\cite{Huang_2012, Jaschik_2019, Ahn_2021}. Following the theoretical prediction, it has already been confirmed experimentally that the hole density can be systematically adjusted by doping with oxygen (O)~\cite{Storm_2021_O} and Se~\cite{Storm_2021_Se}. Moreover, electrodes made of S-doped CuI have been recently fabricated~\cite{Ahn_2022}. These works have shown that p-doping of CuI is possible and in case of the S-electrode increased hole concentrations as well as figure of merit were reported~\cite{Ahn_2022}. Still, as mentioned above, high hole densities were reported to be hindered for some applications~\cite{AoLiu_2021}. Therefore further theoretical studies can be particularly useful to gain a deeper understanding of the modification of electronic properties upon doping and help to identify optimization strategies for a manipulation of a wide range regarding electronic properties. 

In this work we build on promising results and open challenges, and accurately characterize electronic and optical properties of CuI doped with O, S, Se and tellurium (Te). In this way we provide a broad comparison between doping characteristics of the different substitutions considered for p-type doping of CuI. To this end we apply density functional theory using exchange-correlation functionals that has proven to be accurate to characterize the electronic structure of semiconductors\cite{Borlido_2019} and doped crystals~\cite{Zhang_1998,Zhang_2001,Basera_2019, Chen_2019,Yu_2021,Merabet_2022}. We analyze the effect of impurity states on key properties, namely the hole effective mass and the band gap, that determine carrier mobilities and transparency. Our aim is to identify the best p-type dopant for zincblende CuI thin films. 

\begin{figure}
\begin{center}
\includegraphics[width=4cm]{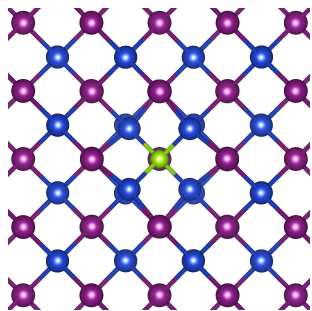} \hfill
\caption{View along the [100] axis of the doped crystal of CuI. Blue balls represent Cu atoms, violet balls symbolize I atoms, and the green ball is a chalcogen dopant. The image was produced using \textsc{Vesta}~\cite{momma_izumi_2011}.}
\label{Bild_SuperzelleDoping}
\end{center}
\end{figure}

\section*{Methods}

All calculations of doped supercells were performed in the framework of density functional theory (DFT), using the projector augmented wave method~\cite{Bloechl_1994} (PAW) as implemented in the Vienna {\it ab initio} simulation package~\cite{Kresse_1996, kresse_joubert_1999} (VASP). The 4$s$ and 3$d$ electrons of Cu, the 5$s$ and 5$p$ electrons of I, as well as the shallow $s$ and $p$ electrons of the chalcogen atoms were treated explicitly as valence electrons in PAW pseudopotentials optimized for GW. A plane-wave basis set with a cutoff energy of 700~eV was used. We approximated the exchange-correlation potential with the Perdew-Burke-Ernzerhof (PBE) functional~\cite{Perdew_1996} and the PBE0 hybrid functional~\cite{PBE0_1996}. To obtain the effective masses an even polynomial of 6$^{th}$~order, $E(k) = ak^6 + bk^4 + ck^2 + d$, was fitted in a k-region around the $\Gamma$-point corresponding to an energy width of 26~meV (i.e., a temperature $T=300$~K). The effective mass was then extracted from the value of $c$ through the relation $E = \frac{\hbar^2k^2}{2m}$. The PBE Kohn-Sham band structure already yields reliable effective masses, but it severely underestimates the size of the band gap. For this reason the hybrid functional PBE0~\cite{PBE0_1996} was applied to correct the band energies and obtain optical spectra. The calculations of the density of states (DOS) were performed with a Gaussian smearing of 0.1~eV. All dielectric functions were obtained in the independent-particle approximation, applying a Lorentzian broadening of 0.1~eV. To describe doping we employed the 64 atom (2$\times$2$\times$2) supercells used in Ref.~\cite{Grauzinyte_2019}, depicted in Fig~\ref{Bild_SuperzelleDoping} together with a 3$\times$3$\times$3 k-point grid. Convergence studies of the supercell size were performed in Ref.\cite{Grauzinyte_2019}. In the Supporting Information we show that the selected supercell size is sufficient also for our purposes, presenting for comparison calculations of the DOS for a 3x3x3 cell (216 atoms) in the case of Se doping. This proves that the overall difference in the DOS is due to the different position of the Fermi level with respect to the valence band maximum, as a consequence of the decreased impurity concentration for the larger supercell. After aligning the Fermi energies, we verify extremely similar DOS. We decide  therefore to favor computational efficiency, using 64 atom supercells together with {\it a posteriori} control of the Fermi energy position through additional charge doping.

\section*{Doping stability}

As outlined in the introduction, the work presented in this manuscript builds on the study performed by some of us in Ref.~\cite{Grauzinyte_2019}. Since the question of doping stability is important, in this section we summarize the results of that study and lay in this way solid foundations for present calculations. In Ref.~\cite{Grauzinyte_2019} we scanned through the periodic table to find suitable p-type dopants for zincblende CuI, taking into account doping both on the Cu and I site. Besides thermodynamic stability, possible transitions into hole- or electron-generating charge states were evaluated to identify suitability for charge carrier population of valence or conduction bands. The formation energy was obtained evaluating
\begin{align}
    E_{D_x}^{F} = E_{D_x}^{q} - E_\mathrm{CuI} - \sum_i n_i \left[ \mu_i + \Delta\mu_i\right] + q\left[ \epsilon_{VBM} + \Delta\epsilon_F\right] + E_{cor} 
\end{align}
where $E_{D_x}^{F}$ is the formation energy for doping with the element $x$, $E_{D_x}^{q}$ the energy of the corresponding supercell in the charge state $q$, $E_\mathrm{CuI}$ the energy of the pristine CuI-supercell, $n_i$ the number of atoms of the species $i$ added to or removed from the crystal and $\mu_i$ the chemical potential of the species $i$. $E_{cor}$ are corrections arising from the supercell approach (see Ref.~\cite{Freysoldt_2014}.) As a conclusion of those calculations, p-type doping was predicted only for chalcogen substitutions on the I-site. Hole concentrations in the range 10$^6$~cm$^{-3}$ - 10$^{19}$~cm$^{-3}$ were anticipated in a temperature interval of 300~K - 600~K for the different elements. More details can be found in Ref.~\cite{Grauzinyte_2019}. A precise characterization of electronic band structures and optical spectra of chalcogen-doped CuI was however beyond the scope of that work. Meanwhile, experiments have confirmed the possibility to realize p-type doping with O~\cite{Storm_2021_O},
Se ~\cite{Storm_2021_Se} an S~\cite{Ahn_2022}, but have identified some difficulties in the performance control that call for a deeper analysis of the underlying electronic properties.

\section*{Effects of doping on the electronic band structure}

\begin{figure}[h]
\begin{center}
\includegraphics[height=5.5cm]{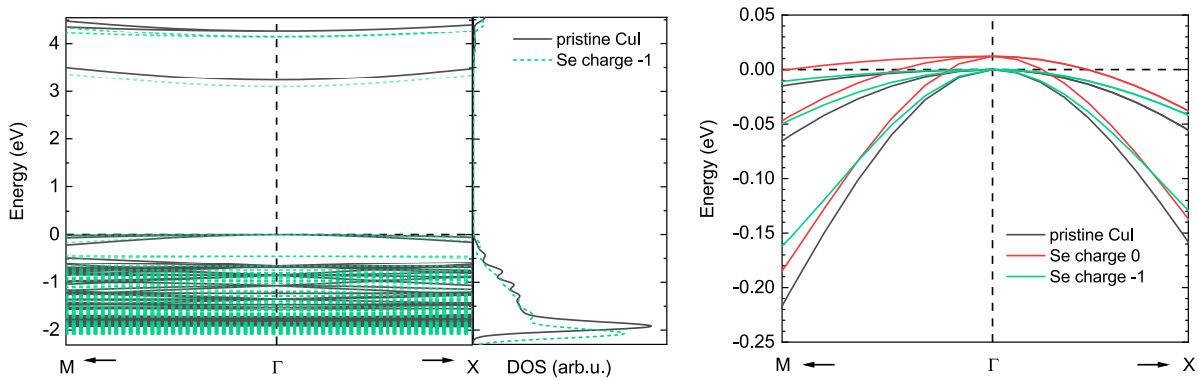} \hfill
\caption{Left panel: Band structure of pristine and Se doped CuI in the hole-generating charge state -1 together with the comparison of the DOS of pristine and doped CuI. Right panel: Zoom-in view of the top valence band: band structure of charged and uncharged Se doped CuI compared with pristine CuI.}
\label{Fig_Bandstructures}
\end{center}
\end{figure}

\begin{figure*}[!htbp]
\begin{center}
\includegraphics[height=7cm]{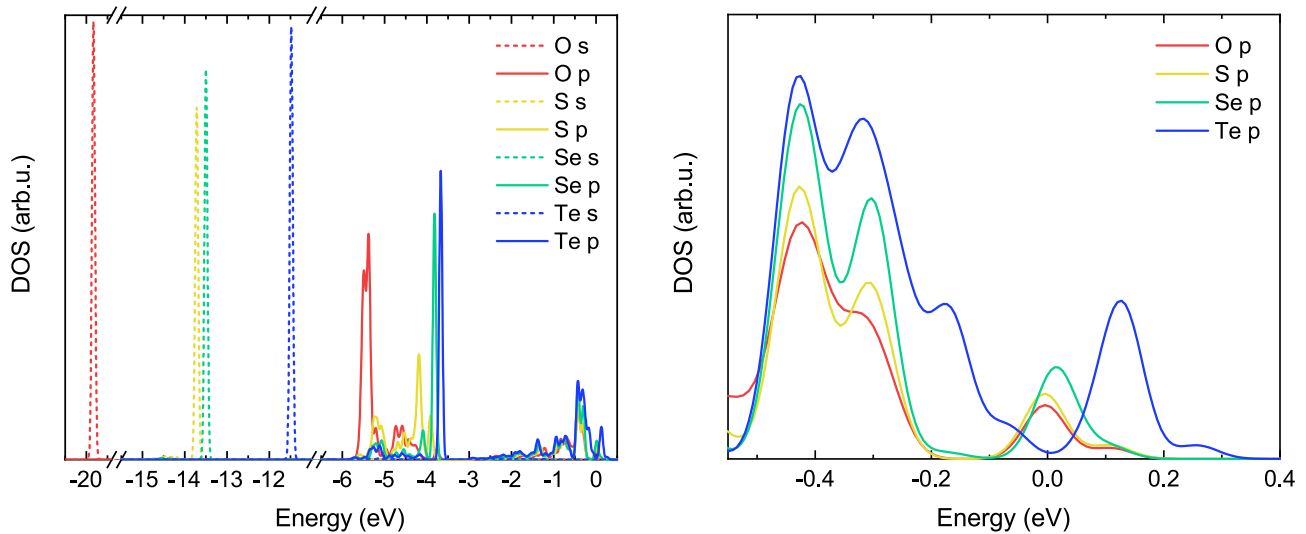} \hfill
\caption{Projected DOS on the impurity atoms for the  doped CuI systems (neutral impurities) calculated with PBE0. On the left panel a wide energy range is shown, on the right panel we focus on a smaller interval around the VBM.}
\label{Fig_DosIon_Chalcogens}
\end{center}
\end{figure*}

Pristine CuI is a natural p-type large-band gap semiconductor and Cu vacancies are recognized as responsible for its intrinsic conductivity~\cite{Grauzinyte_2019}. A possible path to enhance p-type conductivity is to increase hole concentration without inducing a band gap shrinkage. This can be achieved only if defects produce hole-generating charge states (i.e., with impurity charge q~=~-1) located close to the valence band maximum (VBM) and if no additional levels localized on the defect are introduced in the band gap. In this way holes are released into the valence band while the impurities capture electrons. The amount of energy necessary to ionize the defect can be obtained by analyzing the corresponding thermodynamic transition energies. This evaluation was performed by some of us in Ref.~\cite{Grauzinyte_2019}, where all chemical elements were considered as impurity substitutions on either Cu or I sites of the zincblende crystal. In that study chalcogen substitutions on the I site were identified as the only promising p-type dopant, as they display thermodynamic transition levels to a -1 charge state at less than 300~meV above the VBM. To complement the exhaustive thermodynamic investigation of the stability of dopants in CuI, we verified that the supercells that include impurities are dynamically stable by means of phonon calculations. We remark that, following theoretical predictions, successful experimental realisations of Se-doped~\cite{Storm_2021_Se}, O-doped~\cite{Storm_2021_O} and S-doped~\cite{Ahn_2022} CuI have been reported. 

\begin{figure}[h]
\begin{center}
\includegraphics[height=7cm]{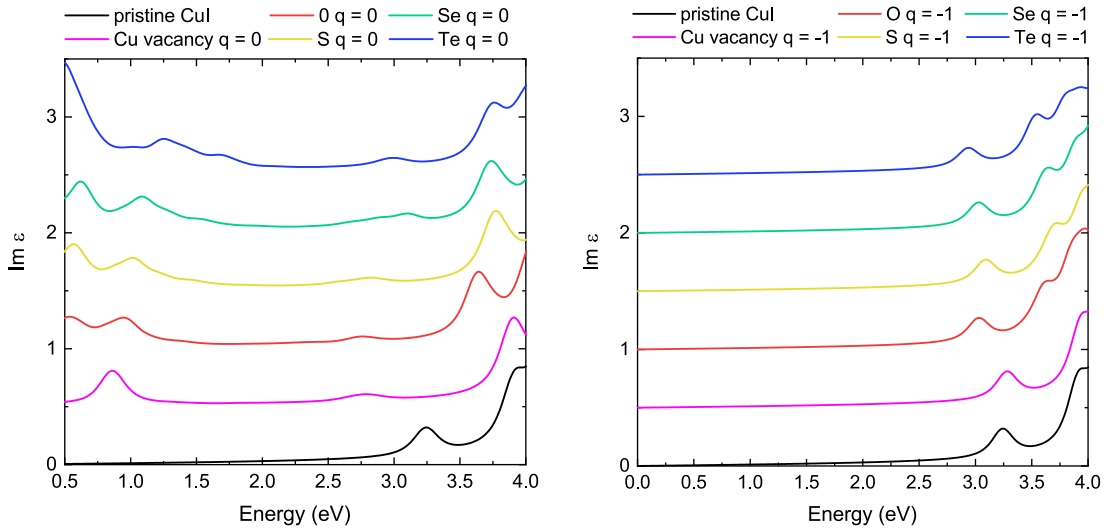} \hfill
\caption{Calculated optical absorption (i.e., imaginary part of the dielectric function) of doped CuI compared with the absorption spectrum of pristine CuI and Cu-deficient CuI. We consider here neutral supercells and the average of the dielectric-function components for orthogonal polarization directions. The spectra have been shifted vertically by steps of 0.5 for clarity. Left: uncharged defect, right: charged defect.}
\label{Fig_Spectrum_Chalcogene}
\end{center}
\end{figure}

As an exemplification of the effects of doping on the electronic energy levels, we show in Fig.~\ref{Fig_Bandstructures} the Kohn-Sham band structure of CuI and Se-doped CuI calculated in the PBE approximation, focusing on the highest valence band. The band structures of CuI doped with the other chalcogen elements are shown in the SI. We remark that, despite the usual underestimation of the band gap, band dispersion (and therefore hole effective masses) are accurately reproduced using this density functional~\cite{seifert2021layered,Grauzinyte_2019}. We have applied therefore a scissor operator of 2.1~eV to the PBE conduction bands shown in Fig~\ref{Fig_Spectrum_Chalcogene} to match the CuI band gap calculated with the PBE0 functional. We remark that we have proved in Ref.~\cite{seifert2021layered} that the PBE0 functional gives a band gap in excellent agreement with experiment, provided that corrections due to spin-orbit interaction are properly accounted for. Here the spin-orbit coupling is neglected, but its band gap renormalization is properly included in the scissor operator. A detailed analysis of effective masses of CuI in the $\gamma$ and $\beta$ phases showed that the effective masses are negligibly affected by SOC.~\cite{seifert2021layered}. Nevertheless, we included in the supporting information calculations on the effect of spin-orbit coupling on the energy levels of all considered doped supercells. We could conclude that the SOC causes a reduction of the band gap of about 0.1-0.2\,eV for all systems, beside inducing band splittings, however the band width close to the VBM is unchanged, pointing to the fact that the effective masses are not significantly affected. 

We can see in Fig.~\ref{Fig_Bandstructures} that a Se defect does not induce deep states in the gap of pristine CuI. This is true for neutral and negatively charged defects. However, the band gap of the doped system is slightly reduced. Also the dispersion of the bands around the VBM remains very similar for doped and undoped system, hinting to a preserved small effective mass. Around 2~eV below the VBM, the width of the bands of the doped system enlarges. This is visible also in the plot of the density of states (DOS) in the right panel of Fig.~\ref{Fig_Bandstructures}. The differences are nevertheless minimal, and we can expect that the presence of such impurities should not compromise optical transparency or hole mobility.

In Fig.~\ref{Fig_Bandstructures} we zoom on the valence band maximum and we show, beside the energy levels of pristine CuI and Se-doped CuI in the negative charge state, also the band structure of the uncharged supercell with a Se impurity. One can see that the effect of charging the impurity leads only to a shift of the Fermi energy. In fact, the Fermi level lies well inside the valence band for the uncharged defect, as a consequence of the high doping concentration in the 64-atom supercell. For the negatively charged Se impurity the chemical potential is instead brought back to the VBM.

We used the example of Se doping to exemplify our findings, but we can observe a similar behavior when CuI is doped with any of chalcogen element. Only Te shows some minor differences as its defect states spread deeper into the gap (as visible in the right panel of Fig.~\ref{Fig_DosIon_Chalcogens}).
The similarities in the DOS for this family of dopants are also reflected in the optical spectra, as we will discuss later.

After this qualitative analysis of the effect of doping on the band structures, we consider more in detail the changes of effective masses of electrons and holes close to the band-gap edges. The values are listed in Table~\ref{Table_me}. 

\begin{table}[ht]
\centering
  \begin{tabular}{llll}
     \hline
           & cb & lh & hh \\
     \hline
pristine CuI & 0.15 & 0.22 & 0.75\\
O q = 0 & 0.16 & 0.24 & 0.95 \\
O q = -1 & 0.16 & 0.38 & 1.55 \\
S q = 0 & 0.15 & 0.23 & 0.84 \\
S q = -1 & 0.15 & 0.25 & 0.97 \\
Se q = 0 & 0.15 & 0.22 & 0.82 \\
Se q = -1 & 0.15 & 0.25 & 0.96 \\ 
Te q = 0 & 0.08 & 0.18 & 0.81 \\
Te q = -1  & 0.15  & 0.24 & 0.93 \\
\hline
  \end{tabular}
   \caption{Comparison of the PBE effective masses of the conduction band (cb), the light-hole band (lh) and the heavy hole band (hh). All effective masses were evaluated from the $\Gamma$-point in direction to the X-point. The results are given in units of the electron mass.}
  \label{Table_me}
\end{table}

One can see that electron and hole effective masses are not modified significantly in the doped systems, and that doped supercells have always a slightly larger value of the effective mass in the negative charge state. Only the charged state of O doped CuI gives a significantly larger heavy-hole mass, but also in this case the light-hole mass remains clearly smaller than the electron mass, indicating that we can still expect a good hole mobility. That larger effective mass for O can be linked to the higher value of electronegativity in comparison with the other doping elements. In fact the larger electronegativity of O than iodine is presented as a reason why CuI has more delocalized holes at the VBM than other Cu oxides \cite{Yamada_2016, Hu_2020} that are also considered for applications as p-type transparent conductors \cite{AoLiu_2021}.

\begin{figure}[h]
\begin{center}
\includegraphics[width=8.5cm]{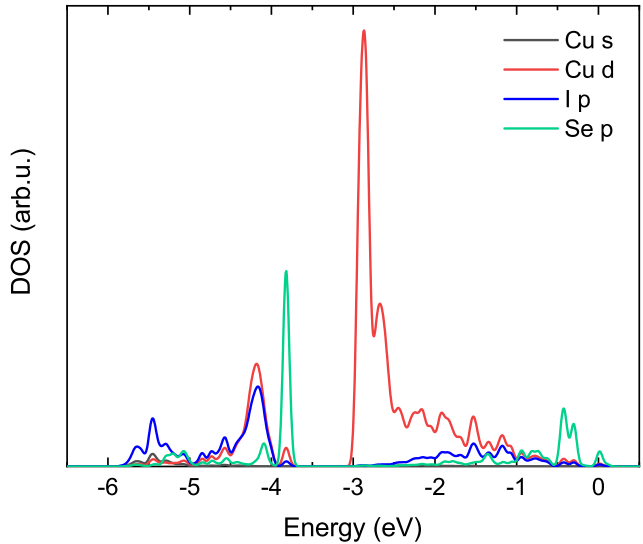} \hfill
\caption{Projected DOS of Se doped CuI (neutral impurity) calculated with the PBE0 hybrid functional. For readability, the contribution of the single Se atom as well as the I atoms were rescaled to take the different number of atoms in the supercell into account.}
\label{Fig_DOS_SeDoped}
\end{center}
\end{figure}

We move now to a deeper analysis of the character of the bands close to the VBM, focusing in particular on the contribution of the dopant states. We show in Fig.~\ref{Fig_DosIon_Chalcogens} the partial DOS and in Fig.~\ref{Fig_DOS_SeDoped} the DOS projected on all atoms. Also in this case we use the case of Se as an exemplification of the results that we found for all considered impurities. Results for the other dopants can be found in the SI. In this case we used the PBE0 hybrid functional to calculate the DOS, as this approximation gives an excellent agreement with experiments~\cite{Grundmann_2013}, in particular concerning the size of the band gap~\cite{seifert2021layered,Grauzinyte_2019}, and it is more accurate to determine the position of the localized valence states and the $p$-$d$ hybridization at the top valence. 

In the left panel of Fig.~\ref{Fig_DosIon_Chalcogens} we observe at lower energies sharp peaks due to the localized $s$ orbitals of the dopants. These states in the valence are strongly localized on the impurities and they are energetically ordered according to the electronegativity of the dopant: O has the highest electronegativity, i.e. the strongest ability to attract electrons, and its $s$ level reaches the lowest energy of about -20\,eV. The $p$ orbitals of the dopants are found closer to the VBM.  If we zoom into the energy interval close to the Fermi energy, shown in the right panel of Fig.~\ref{Fig_DosIon_Chalcogens}, we can see that all dopant elements yield peaks of the DOS of the neutral supercell slightly above the VBM. These shallow defect states are those that enable p-type doping when holes are transferred to the valence band. Te has the smallest electronegativity and by far the most contribution of the DOS in this region up to around 0.3\, eV above the Fermi-level. 

In Fig.~\ref{Fig_DOS_SeDoped} we consider the contribution of all atoms to the DOS of the Se doped system in the uncharged state. To make visible the individual states, we multiplied the contributions of Se by 32, the ones of I by 32 divided by 31, while leaving the Cu DOS unchanged, therefore accounting for the different numbers of atoms in the unit cell. We can observe that the top of the valence band is made of Se and I $p$ states hybridized with Cu $d$ states and we see the gap induced by $p$-$d$ repulsion between bonding and antibonding hybridized bands. It is clearly visible that at the VBM the contribution of Se states dominates, making this dopant a shallow acceptor.

\begin{figure}[h]
\begin{center}
\includegraphics[height=10cm]{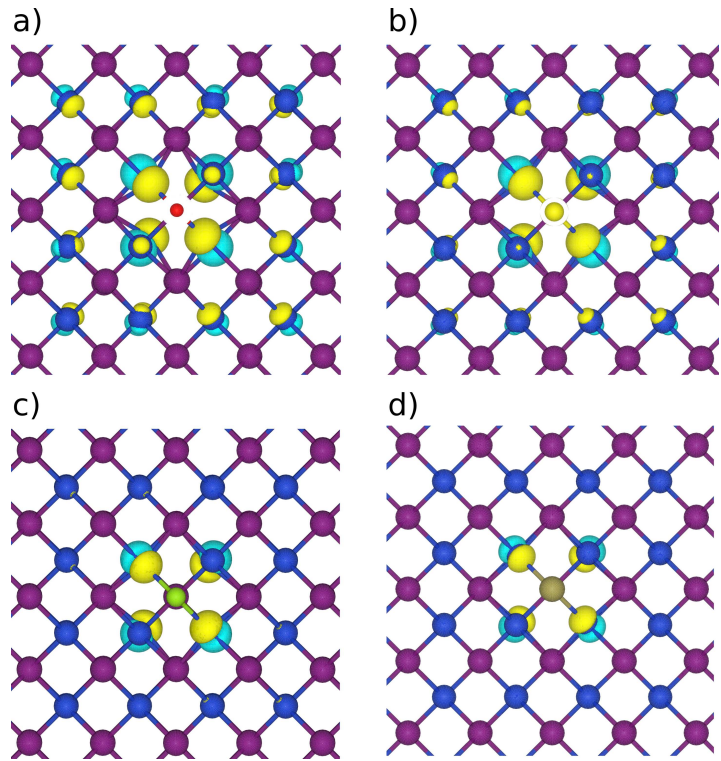} \hfill
\caption{Difference of electron density between the uncharged doped CuI cell and pristine CuI. We consider substitutional doping on the I site with a) oxygen, b) sulfur, c) selenium and d) tellurium. Yellow colors indicate positive charge differences, light blue negative charge differences. The charge density was calculated with PBE-DFT and the images were made using VESTA~\cite{momma_izumi_2011}.}
\label{Fig_CHGTransfer}
\end{center}
\end{figure}

In Fig.~\ref{Fig_CHGTransfer} we analyse charge transfer in the proximity of the impurity, considering the difference of electron-density between the supercells with an uncharged impurity on the I site and pristine CuI. We observe for all considered dopants a redestribution of charge (a positive charge variation is indicated in yellow, while negative charge variations are light blue) around the neighbouring Cu atoms. For O and S impurities the perturbation concerns not only the nearest neighbours, but it has a considerable longer range.

\section*{Effects of doping on the optical properties}

We calculated the dielectric functions of chalcogen-doped CuI in the independent-particle transition picture, neglecting excitonic effects. To this aim we used the band structures obtained with the hybrid functional PBE0 that correctly reproduces the band gap of CuI. 

In Fig.~\ref{Fig_Spectrum_Chalcogene} we display the imaginary part of the dielectric function, i.e. the optical absorption spectrum, of pristine CuI and the doped supercells in the neutral charge state. We included for comparison the absorption spectrum of CuI with a neutral Cu vacancy in the supercell. 
As expected, all systems except pristine CuI have a notably high absorption in the infrared that extends to the visible range, due to transitions from occupied valence states to empty states close to the VBM. This applies especially for Te doping, and it is in agreement with the observations that we made concerning Fig~\ref{Fig_DosIon_Chalcogens}. We can further observe that the doped systems are less transparent than CuI with Cu vacancies.

The situation is different when we consider a negatively charged supercell, with Fermi energy closer to the top valence as expected also in case of a more realistic lower dopant concentration, as it is shown in Fig.~\ref{Fig_Spectrum_Chalcogene}. In this case the Fermi level is pushed back to the VBM and low-lying transitions are suppressed. Transparency is ensured up to around 2.5~eV for all systems. The S doped compound has the best transparency in the UV, loosing there only to the Cu vacancy. Keeping the higher charged defect concentration of S doped CuI shown in \cite{Grauzinyte_2019}, doping with S still seems superior to the natural occurring Cu vacancies even though the transparency of the doped system is slightly worse in the UV. Of course, the actual position of the Fermi level depends on the doping concentration, that is in our case fixed by the size of the supercell. Se doping is probably the most promising as it should induce a smaller geometry deformation (O and S atoms are much smaller than I) and produce very shallow defect states with small effective masses. 
On the other hand, compensating doping on the Cu site can be envisaged if too high hole-densities cause undesirable effects (such as the poor current modulation reported for CuI-based thin film transistors \cite{AoLiu_2021}). Several potential substitutional dopants on the Cu site were identified in Ref.\cite{Grauzinyte_2019} that can counteract p-type doping, even if the possibility of ambipolar doping was excluded by calculations. 
Additional optical spectra (real part of the dielectric function, real and imaginary part of the refractive index and reflectivity) can be found in Fig.~S4 of the supporting information.

\section*{Conclusion} 
We present here an \textit{ab initio} study on the effects substitutional doping on the I site with chalcogenes to manipulate electronic and optical properties of CuI. The latter is a promising material for p-type transparent contacts, but technological breakthroughs require a conductivity enhancement in this semiconductor that does not deteriorate transparency and hole mobility. 

The choice of this specific elements as dopants is based on the results of our recent high-throughput study~\cite{Grauzinyte_2019} that selected chalcogens as the best candidates for effective $p$-doping. A precise characterization of electronic and optical properties of doped CuI is further called for by the very recent experimental realization of chalcogen doping. 

We find that the position and characteristics of impurity energy levels, as well as the values of electron and hole effective masses are very similar across this family of dopants. Our calculations show that sulfur guarantees the best transparency, and both sulfur and selenium lead to significant increase of charge carrier concentrations. Also in view of the absence of charge localization at the valence band maximum, we conclude that substitutions of iodine with sulfur and selenium is the most promising way to enhance hole concentration without deteriorating carrier mobility and transparency. Our results suggest that further experimental investigation of chalcogen doping of CuI to optimize its performance as transparent conductor is certainly worthwhile being pursued. 

\section*{Acknowledgments}
We thank José Flores-Livas and Migl\.{e} Grau\u{z}inyt\.{e} for fruitful discussions and for providing the supercells of doped systems. This work received funding from the Deutsche Forschungsgemeinschaft (DFG) through the research unit FOR~2857 and the project BO~4280/8-1. Computational resources were provided by the Leibniz Supercomputing Centre on SuperMUC (project pn68le).

\section*{Supporting Information}

In the Supporting Information we include additional band structures and optical spectra complementing the information of the main manuscript. Furthermore we discuss the influence of the supercell size and spin-orbit coupling on the quantities discussed in the main manuscript.

\section*{Author contributions statement}
S.B. initiated the study. M.S. did the DFT-calculations. M.S, M.A.L.M. and S.B. analyzed and discussed the data. All authors wrote and reviewed the manuscript.

\section*{Notes}
The authors declare no competing financial interests.

\section*{Data availability}
The data that support the findings of this study are available within this article and its supplementary material or are available from the corresponding authors upon reasonable request.

\bibliography{CuIDoping}

\end{document}